\documentclass{article}

\usepackage[utf8]{inputenc}
\usepackage[T1]{fontenc}
\usepackage{arxiv}
\usepackage{amsmath,amsfonts,amssymb}
\usepackage{graphicx}
\usepackage{float}
\usepackage{booktabs}
\usepackage[numbers]{natbib}
\usepackage[colorlinks=true, allcolors=blue]{hyperref}
\usepackage{authblk}
\usepackage{siunitx}

\title{Benchmarking Open-Source FDK Against Commercial and Iterative Reconstruction Methods for Preclinical Micro-CBCT}
\renewcommand{\shorttitle}{FDK vs Iterative Reconstruction for Micro-CBCT}

\author[a]{Falk L. Wiegmann}
\author[a,b]{Nancy L. Ford\thanks{Corresponding author: nlford@dentistry.ubc.ca}}
\affil[a]{Department of Physics and Astronomy, The University of British Columbia, Vancouver BC, Canada}
\affil[b]{Department of Oral Biological and Medical Sciences, The University of British Columbia, Vancouver BC, Canada}

\begin{document}

\maketitle

\begin{abstract}
Preclinical micro-CT reconstruction involves large projection sizes and volumes that make iterative methods costly---$5\times$ to $50\times$ slower than analytic alternatives on modern GPUs. Whether this cost is justified depends on the imaging task, yet head-to-head comparisons using task-based metrics on identical preclinical data are lacking. We benchmark four reconstruction methods on identical acquisitions from an eXplore CT~120 micro-CT scanner (Trifoil Imaging, USA): an open-source Feldkamp--Davis--Kress (FDK) pipeline, the proprietary vendor software, and two iterative toolboxes at default settings---ASTRA SIRT and TIGRE OS-SART. Using the modulation transfer function (MTF), noise power spectrum (NPS), and non-prewhitening detectability index (NPW~$d'$), we show that single-metric rankings are misleading: the vendor software achieves the highest spatial resolution (MTF$_{10} = \SI{2.96}{lp/mm}$) but fails to reach the Rose criterion ($d' = 3$) for \SI{100}{HU} contrast objects on a half-scan acquisition. ASTRA SIRT, at $5\times$ the computation time of FDK, provides the best low-contrast detectability, while TIGRE OS-SART at $50\times$ the cost offers no additional benefit and exhibits instability across scan protocols. For high-contrast tasks, all methods perform comparably. We release our FDK pipeline as open-source software, providing a fast, transparent, and integrable reconstruction tool for the preclinical micro-CT community.
\end{abstract}


%% ============================================================
\section{Introduction}\label{sec:introduction}
%% ============================================================

Preclinical micro-CT is a high-throughput modality: longitudinal studies routinely require reconstructing dozens of scans per cohort, and respiratory- or cardiac-gated acquisitions multiply that count further~\cite{ford_respiratory_2025}. The computational burden of reconstruction is particularly acute in micro-CT compared to clinical CT, because the high spatial resolution demands both large projection sizes (here $3500 \times 2300$~pixels) and correspondingly large reconstruction volumes (over $10^9$~voxels), each requiring approximately \SI{19}{GiB} of GPU memory. On a modern GPU (NVIDIA H100), the Feldkamp--Davis--Kress (FDK) algorithm~\cite{feldkamp_fdk_1984} reconstructs a full volume in approximately \SI{4.5}{min}. The Simultaneous Iterative Reconstruction Technique (SIRT)~\cite{gilbert_sirt_1972}, as implemented in ASTRA~\cite{van_aarle_astra_2016}, requires \SI{23.8}{min} for 100~iterations---roughly $5\times$ longer---while Ordered-Subset Simultaneous Algebraic Reconstruction (OS-SART)~\cite{andersen_sart_1984}, as implemented in TIGRE~\cite{biguri_tigre_2016}, requires \SI{226}{min}, more than $50\times$ longer. For a 20-animal study, these differences translate to 1.5~hours, 8~hours, and over 3~days of reconstruction time, respectively.

Despite this cost disparity, iterative methods are widely regarded as producing superior image quality due to their ability to suppress noise without the bandwidth-limiting filters required by analytic methods. However, most published comparisons between analytic and iterative reconstruction in preclinical micro-CT report only single metrics such as the modulation transfer function (MTF) or noise magnitude, which do not capture how resolution and noise jointly affect the ability to detect features of interest. Task-based metrics such as the non-prewhitening (NPW) detectability index $d'$~\cite{aapm_tg233_2019} provide a more complete assessment, yet head-to-head comparisons using these metrics on identical preclinical data are lacking.

A further barrier is transparency. Most preclinical labs rely on vendor-supplied reconstruction software that operates as a black box: filter kernels, correction algorithms, and calibration procedures are undocumented and cannot be modified. This opacity prevents users from diagnosing image quality limitations or adapting reconstruction parameters to their specific imaging task. It also precludes integration into emerging workflows such as dual-domain deep neural network training~\cite{zhang_sparse_2022}, where differentiable forward and inverse operators must be embedded directly in the training pipeline, or sinogram-domain processing where access to intermediate representations is essential.

In this work, we benchmark four reconstruction methods on identical preclinical micro-CT data: our open-source FDK implementation, the proprietary GE Healthcare (vendor) software bundled with the eXplore CT~120 scanner, and two widely used iterative toolboxes at their default settings---ASTRA~\cite{van_aarle_astra_2016} (SIRT) and TIGRE~\cite{biguri_tigre_2016} (OS-SART). We evaluate each method using the MTF, noise power spectrum (NPS), and NPW~$d'$~\cite{aapm_tg233_2019,icru_ct_2012} on a quality assurance phantom, supplement the quantitative analysis with a visual comparison on an in vivo mouse lung scan, and assess robustness across half-scan and full-scan acquisition protocols. Our results show that single-metric rankings are misleading: the method with the highest spatial resolution fails to detect low-contrast features, while the $50\times$ slower iterative method provides only marginal detectability gains over the $5\times$ slower alternative. We release our FDK pipeline as open-source software to provide the preclinical community with a fast, transparent, and integrable reconstruction tool.

%% ============================================================
\section{Methods}\label{sec:methods}
%% ============================================================

\subsection{Scanner and Acquisition}\label{sec:scanner}

All data were acquired on the eXplore CT~120 micro-CT scanner (Trifoil Imaging, USA) at \SI{80}{kVp}, \SI{40}{mA}, with \SI{16}{ms} exposure per frame and no frame averaging. The flat-panel detector comprises $3500 \times 2300$ pixels. Table~\ref{tab:scan_geometry} summarises the geometric parameters for each dataset.

\begin{table}[H]
\centering
\caption{Scanner geometry and scan protocols. SDD: source-to-detector distance. COR: centre of rotation.}
\label{tab:scan_geometry}
\small
\begin{tabular}{lcc}
\toprule
Parameter & Phantom (1988/1989) & Mouse (1510) \\
\midrule
Detector pixel size (mm) & 0.0284 & 0.0283 \\
SDD (mm) & 451.5 & 448.8 \\
Source--isocentre (mm) & 396.4 & 396.7 \\
COR pixel & 1745.0 & 1758.5 \\
Central slice pixel & 1121.0 & 1137.5 \\
\midrule
Scan 1988 & Half-scan: 220 proj, \ang{0.877}/proj, \ang{193} & --- \\
Scan 1989 & Full-scan: 440 proj, \ang{0.818}/proj, \ang{360} & --- \\
Scan 1510 & --- & Half-scan: 220 proj, \ang{193} \\
\bottomrule
\end{tabular}
\end{table}

\subsection{Phantom and Mouse}\label{sec:phantom}

Quantitative image quality was evaluated on the mCTP~610 quality assurance phantom (Shelley Medical Imaging Technologies, Canada), which contains a slanted-edge insert (air/acrylic boundary) for MTF measurement and a homogeneous water region for NPS characterisation. Calibration inserts span $-940$ to \SI{1460}{HU}. Two scans of the phantom were acquired: a half-scan (scan~1988, \ang{193}) and a full-scan (scan~1989, \ang{360}).

Visual comparison was performed on an in vivo mouse lung scan acquired with respiratory gating so that projections correspond to a single phase of the respiratory cycle, avoiding motion blur (half-scan, 220~projections).

\subsection{Reconstruction Methods}\label{sec:recon_configs}

All non-vendor methods were applied to identically preprocessed sinogram data (flat-field correction, beam hardening correction, ring artefact suppression, cone-beam weighting, and Parker weighting~\cite{parker_weights_1982} for short scans) and used a \SI{0.075}{mm} isotropic voxel size. All outputs were converted to Hounsfield units via two-point calibration (air and water references) prior to metric evaluation. The phantom field of view was $93.5 \times 93.5 \times \SI{70}{mm}$ ($1247 \times 1247 \times 933$~voxels); the mouse was reconstructed over a cropped ROI of approximately $24.8 \times 25.9 \times \SI{57.3}{mm}$ ($\sim$$330 \times 345 \times 764$~voxels). Reconstructions were performed on an NVIDIA H100 GPU (\SI{128}{GiB}); each method required approximately \SI{19}{GiB} of VRAM for a full-scan reconstruction at the phantom volume size. Our FDK implementation supports dynamic z-slice chunking, allowing reconstruction on GPUs with less available memory or on CPU, at the expense of longer computation time. Table~\ref{tab:recon_methods} summarises the method-specific parameters.

\begin{table}[H]
\centering
\caption{Reconstruction method parameters. Wall-clock times are for a full-scan phantom reconstruction on an NVIDIA H100 GPU.}
\label{tab:recon_methods}
\small
\begin{tabular}{lp{8.5cm}r}
\toprule
Method & Configuration & Time \\
\midrule
Our FDK & Hamming window; cutoff $= d_a / d_x = 0.0284 / 0.075 \approx 0.378 \times$~Nyquist (matched to magnification). Ramp filter in physical units (mm$^{-1}$) with the standard FDK $1/2$ prefactor; output in linear attenuation (mm$^{-1}$). & \SI{4.5}{min} \\
\addlinespace
Vendor Software & Vendor-proprietary FDK from the eXplore CT~120 software. Voxel size \SI{0.0748}{mm} isotropic. Output in int16 big-endian format with reformatting bins $y{=}3$, $z{=}3$. Internal filter and correction details are undocumented. & --- \\
\addlinespace
ASTRA SIRT~\cite{van_aarle_astra_2016} & astra-toolbox 2.2; \texttt{SIRT3D\_CUDA}; 100~iterations; non-negativity constraint (min~$= 0$); no upper constraint. & \SI{23.8}{min} \\
\addlinespace
TIGRE OS-SART~\cite{biguri_tigre_2016} & TIGRE 2.6; \texttt{ossart}; 100~iterations; 15~projections per subset; relaxation $\lambda = 0.5$ with per-iteration decay $\lambda_\mathrm{red} = 0.97$; no non-negativity constraint. \texttt{offOrigin} forced to $[0,0,0]$ (non-zero values caused a CUDA hang); off-centre ROIs handled via expand-and-crop. & \SI{226}{min} \\
\bottomrule
\end{tabular}
\end{table}

\subsection{Image Quality Metrics}\label{sec:evaluation}

\subsubsection{Modulation Transfer Function}

The MTF was measured from the slanted-edge insert using an ISO~12233-style~\cite{iso_12233_2017} oversampled edge technique with $4\times$ subpixel alignment. The line spread function was obtained by differentiation of the edge response function and fitted with a Gaussian ($\sigma \in [0.5,\; 5] \times$~pixel size); signal was retained within $6 \times$~FWHM and margins zeroed at $12 \times$~FWHM. The MTF was computed as the normalised magnitude of the Fourier transform of the processed LSF.

\subsubsection{Noise Power Spectrum}

The NPS was measured following ICRU Report~87~\cite{icru_ct_2012} from eight $132 \times 132$~pixel ROIs arranged in a ring pattern (\ang{0} to \ang{315} in \ang{45} steps) over 16~axial slices in the homogeneous water region. A third-order polynomial was subtracted from each ROI to remove low-frequency trends. The 2D NPS was computed as $|\mathrm{FFT}|^2$, normalised by $\Delta x^2 / (\pi r^2 N_\mathrm{ROI})$, and radially averaged. The integrated NPS ($\int$NPS) was obtained by 2D Simpson integration.

\subsubsection{Detectability Index}

Task-based detectability was quantified using the non-prewhitening matched filter (NPW) observer model~\cite{aapm_tg233_2019}. The task function was a circular disc of contrast $\Delta C$ and radius $R$: $W(f) = \Delta C \, \pi R^2 \, 2J_1(2\pi R f) / (2\pi R f)$. The detectability index was computed as
\begin{equation}
    d'^2 = \frac{\left[\int_0^{f_\mathrm{max}} |W(f)|^2 \, \mathrm{MTF}^2(f) \, f \, df\right]^2}{\int_0^{f_\mathrm{max}} |W(f)|^2 \, \mathrm{MTF}^2(f) \, \mathrm{NPS}(f) \, f \, df}
\end{equation}
for disc diameters linearly spaced from 0.1 to \SI{5.0}{mm} (80~points) at contrast levels of 100 and \SI{500}{HU}. A noise floor of $10^{-12} \times \max(\mathrm{NPS})$ was applied to prevent division by zero.

\subsection{Use of AI Tools}\label{sec:ai_tools}

Claude (Anthropic) was used to assist with manuscript preparation, including editing text and grammar review, and coding assistance during data analysis. All AI-generated content was reviewed, verified, and revised by the authors, who take full responsibility for the accuracy and integrity of the final manuscript.

\subsection{Ethics Statement}\label{sec:ethics}

No animal experiments were conducted in the present study. The only new data acquired for this work were scans of an mCTP~610 quality assurance phantom (Shelley Medical Imaging Technologies, Canada), which were used exclusively for quantitative evaluation. The mouse micro-CT imaging data used for visual comparison were collected as part of previously published studies~\cite{ford_respiratory_2025,ford_spie_2023}. All animal procedures in those studies were approved by the University of British Columbia Animal Care Committee (Protocol No.\ A21-0060, approved August~31, 2021) and performed in accordance with the ARRIVE guidelines~\cite{percie_du_sert_arrive_2020}. No animals were imaged, handled, or subjected to any procedures as part of this work.

%% ============================================================
\section{Results}\label{sec:results}
%% ============================================================

\subsection{Phantom Image Quality Metrics}\label{sec:quantitative_results}

\begin{figure}[H]
    \begin{center}
    \includegraphics[width=\textwidth]{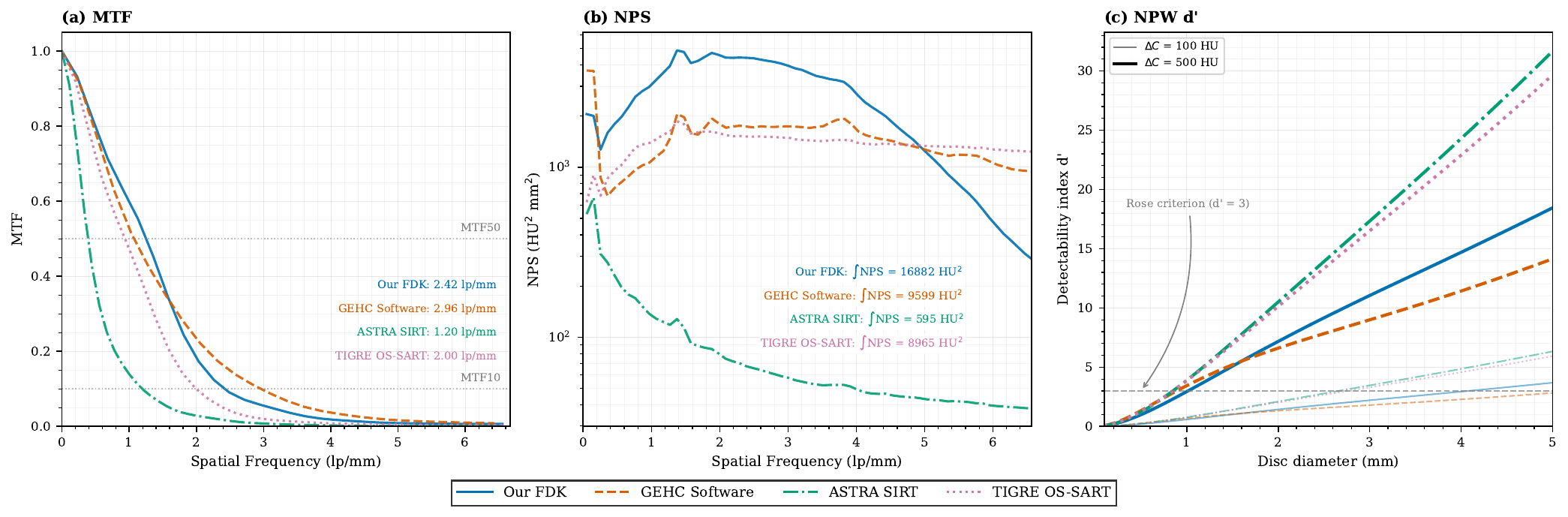}
    \end{center}
    \caption{Image quality metrics for the half-scan (\ang{193}) phantom acquisition. (a)~Modulation transfer function with MTF$_{10}$ values annotated. (b)~Noise power spectrum on a logarithmic scale with integrated NPS values. (c)~NPW detectability index as a function of disc diameter at $\Delta C = 100$ and \SI{500}{HU}; the dashed horizontal line marks the Rose criterion ($d' = 3$).}
    \label{fig:recon_comparison_metrics}
\end{figure}

Figure~\ref{fig:recon_comparison_metrics} presents the MTF, NPS, and NPW~$d'$ measured on the half-scan (\ang{193}) phantom acquisition. The methods exhibited a clear resolution--noise tradeoff. The vendor software achieved the highest MTF$_{10}$ (\SI{2.96}{lp/mm}), followed by our FDK (\SI{2.42}{lp/mm}), TIGRE OS-SART (\SI{2.00}{lp/mm}), and ASTRA SIRT (\SI{1.20}{lp/mm}). Noise showed the inverse ranking: ASTRA SIRT had an integrated NPS of \SI{595}{HU^2}, roughly $28\times$ lower than our FDK (\SI{16882}{HU^2}), with TIGRE OS-SART (\SI{8965}{HU^2}) and the vendor software (\SI{9599}{HU^2}) in between.

The NPW~$d'$ captures this tradeoff in a single task-relevant metric (Table~\ref{tab:rose_half}). At high contrast ($\Delta C = \SI{500}{HU}$), all methods crossed the Rose criterion~\cite{rose_vision_1973} ($d' = 3$) below \SI{1.01}{mm}, with the iterative methods reaching the threshold at \SI{0.85}{mm}. At low contrast ($\Delta C = \SI{100}{HU}$), differences were more pronounced: the iterative methods crossed at 2.66--\SI{2.76}{mm}, our FDK at \SI{4.08}{mm}, and the vendor software never reached $d' = 3$ (maximum $d' = 2.83$), indicating that its elevated low-frequency noise limits low-contrast detectability despite its superior resolution.

\begin{table}[H]
\centering
\caption{Minimum disc diameter (mm) at which each method crosses the Rose criterion ($d' = 3$) on the half-scan phantom. Smaller values indicate better detectability.}
\label{tab:rose_half}
\small
\begin{tabular}{lcc}
\toprule
Method & $\Delta C = \SI{100}{HU}$ & $\Delta C = \SI{500}{HU}$ \\
\midrule
Our FDK       & 4.08 & 1.01 \\
Vendor Software & never ($d'_\mathrm{max} = 2.83$) & 0.89 \\
ASTRA SIRT    & 2.66 & 0.85 \\
TIGRE OS-SART & 2.76 & 0.85 \\
\bottomrule
\end{tabular}
\end{table}

\subsection{Visual Comparison on Mouse Lung}\label{sec:visual_results}

\begin{figure}[H]
    \begin{center}
    \includegraphics[width=\textwidth]{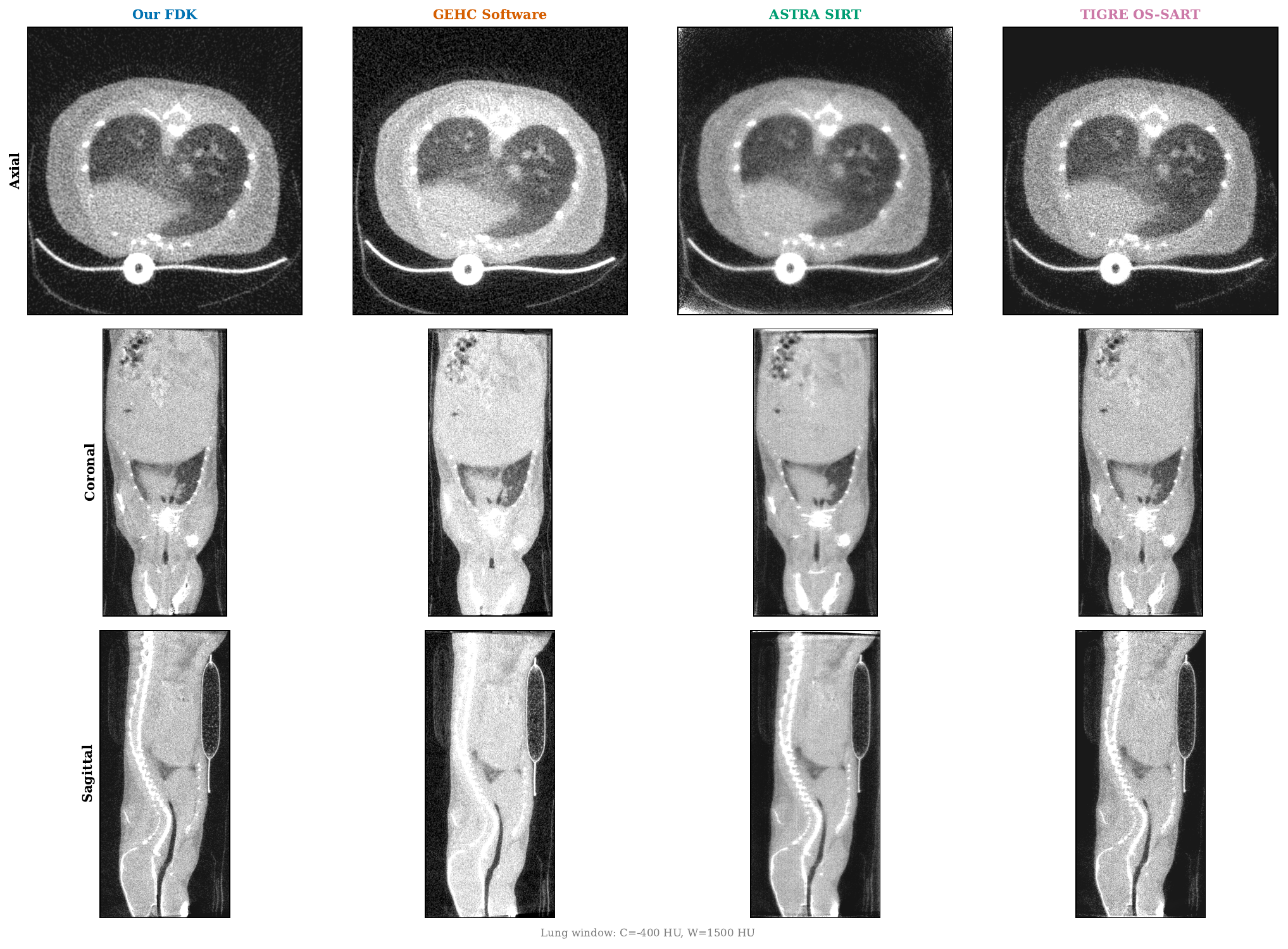}
    \end{center}
    \caption{Mouse lung reconstructions displayed in a lung window (C$= -400$~HU, W$= 1500$~HU). Columns: Our FDK, Vendor Software, ASTRA SIRT, TIGRE OS-SART. Rows: axial, coronal, and sagittal slices. The half-scan (\ang{193}) respiratory-gated acquisition was used.}
    \label{fig:mouse_recon_comparison}
\end{figure}

Figure~\ref{fig:mouse_recon_comparison} shows axial, coronal, and sagittal slices of the mouse lung reconstructed with each method, displayed in a lung window (C$= -400$~HU, W$= 1500$~HU). Our FDK and the vendor software produced images with comparable noise texture and clear depiction of fine anatomical detail such as the rib cage and airways. ASTRA SIRT produced visibly smoother images with reduced noise but with a loss of fine structural detail, consistent with its lower measured MTF. TIGRE OS-SART yielded an intermediate appearance, preserving more structural detail than ASTRA SIRT while showing less noise than the FDK methods. These visual impressions are consistent with the quantitative resolution--noise tradeoffs measured on the phantom (Fig.~\ref{fig:recon_comparison_metrics}).

\newpage

\subsection{Consistency Across Scan Protocols}\label{sec:scan_comparison}

\begin{figure}[H]
    \begin{center}
    \includegraphics[width=\textwidth]{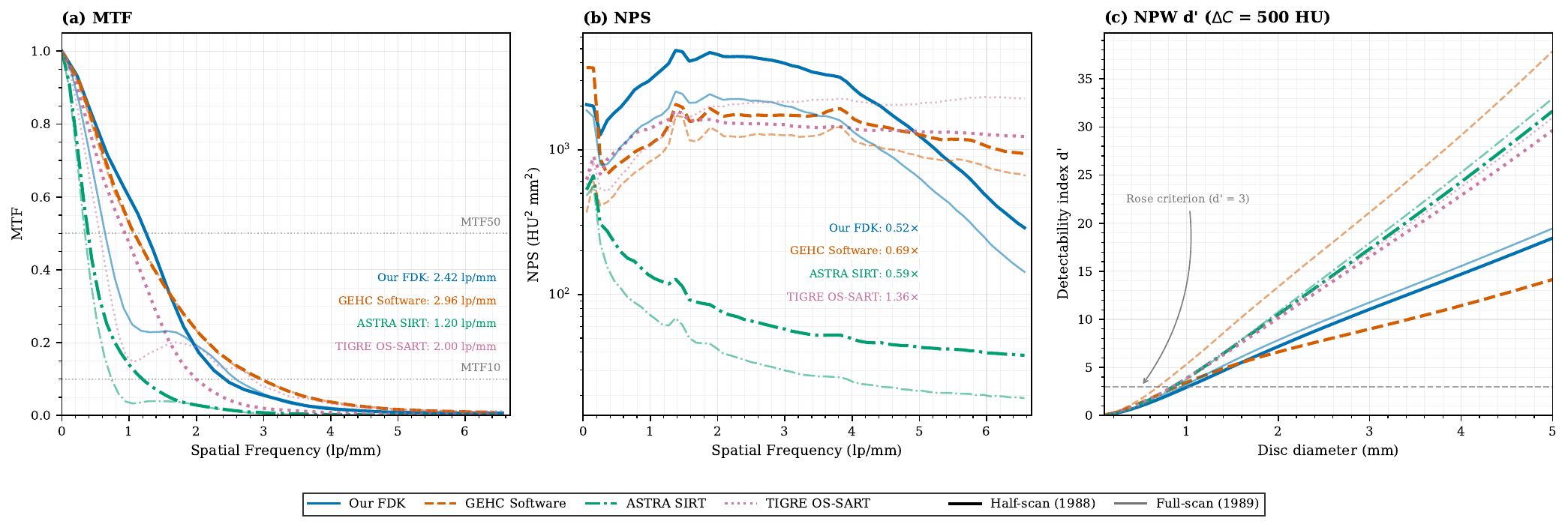}
    \end{center}
    \caption{Comparison of image quality metrics between the half-scan (\ang{193}, bold lines) and full-scan (\ang{360}, thin lines) phantom acquisitions. (a)~MTF curves for both protocols. (b)~NPS with the ratio of integrated NPS (full/half) annotated for each method. (c)~NPW~$d'$ at $\Delta C = \SI{500}{HU}$.}
    \label{fig:scan_comparison_metrics}
\end{figure}

To assess robustness, the phantom was also reconstructed from a full-scan (\ang{360}, 440~projections) acquisition. Figure~\ref{fig:scan_comparison_metrics} overlays the half- and full-scan metrics for each method. The primary difference between protocols was in noise: our FDK, vendor, and ASTRA SIRT all showed reduced NPS with the full scan (ratios of $0.52\times$, $0.69\times$, and $0.59\times$ respectively), consistent with the expected noise reduction from doubling the number of projections. TIGRE OS-SART was the exception, with its integrated NPS increasing by a factor of $1.36\times$ for the full scan, suggesting that its relaxation schedule may not be optimally tuned for the higher projection count.

Table~\ref{tab:rose_comparison} compares the Rose criterion crossings across scan protocols. At $\Delta C = \SI{500}{HU}$, the full scan provided modest improvements for most methods, though ASTRA SIRT showed a slight degradation (0.85 to \SI{0.88}{mm}). At $\Delta C = \SI{100}{HU}$, the vendor software improved from never reaching $d' = 3$ on the half scan to crossing at \SI{2.21}{mm} on the full scan, the largest gain of any method. The iterative methods showed only minor changes, while our FDK improved slightly from 4.08 to \SI{3.86}{mm}.

\begin{table}[H]
\centering
\caption{Rose criterion crossings (disc diameter in mm) for half-scan vs.\ full-scan acquisitions. Smaller values indicate better detectability.}
\label{tab:rose_comparison}
\small
\begin{tabular}{lcccc}
\toprule
 & \multicolumn{2}{c}{$\Delta C = \SI{100}{HU}$} & \multicolumn{2}{c}{$\Delta C = \SI{500}{HU}$} \\
\cmidrule(lr){2-3} \cmidrule(lr){4-5}
Method & Half & Full & Half & Full \\
\midrule
Our FDK       & 4.08 & 3.86 & 1.01 & 0.95 \\
Vendor Software & never & 2.21 & 0.89 & 0.69 \\
ASTRA SIRT    & 2.66 & 2.59 & 0.85 & 0.88 \\
TIGRE OS-SART & 2.76 & 2.69 & 0.85 & 0.85 \\
\bottomrule
\end{tabular}
\end{table}

\subsection{Half-Scan vs.\ Full-Scan Residual Analysis}\label{sec:residual_results}

\begin{figure}[H]
    \begin{center}
    \includegraphics[width=\textwidth]{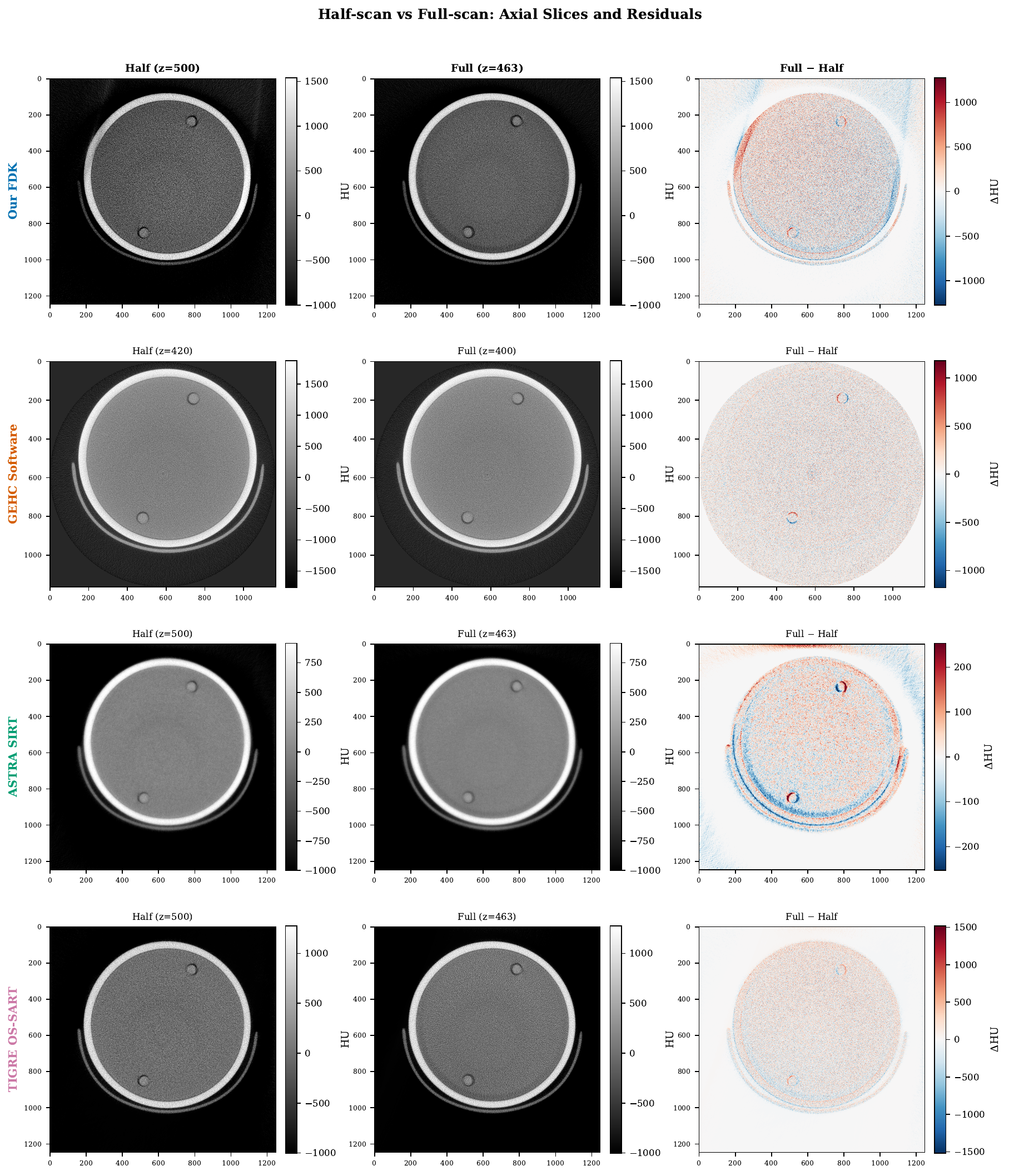}
    \end{center}
    \caption{Axial phantom slices from the half-scan (left column), full-scan (centre column), and their difference (right column, full$-$half) for each reconstruction method. Note the varying $\Delta$HU colour scale ranges across methods: our FDK and vendor show residuals within $\pm$\SI{1000}{HU}, ASTRA SIRT within $\pm$\SI{200}{HU}, and TIGRE OS-SART within $\pm$\SI{1500}{HU}.}
    \label{fig:scan_comparison_residuals}
\end{figure}

Figure~\ref{fig:scan_comparison_residuals} shows the full$-$half residual images for each method. ASTRA SIRT exhibited the smallest residual magnitude ($\pm$\SI{200}{HU}) but with structured ring-like patterns near the calibration inserts, suggesting that iterative convergence differed between the two angular sampling conditions. Our FDK and the vendor software showed larger residuals ($\pm$\SI{1000}{HU}), dominated by noise differences. TIGRE OS-SART showed the largest residuals ($\pm$\SI{1500}{HU}) with pronounced structured differences, consistent with its anomalous NPS increase for the full-scan protocol.

%% ============================================================
\section{Discussion}\label{sec:discussion}
%% ============================================================

The central finding of this work is that single image quality metrics do not predict task performance. The vendor software achieved the highest spatial resolution of any method tested, yet it was the only method that failed to reach the Rose criterion~\cite{rose_vision_1973} at $\Delta C = \SI{100}{HU}$ on the half-scan acquisition. This limitation is invisible to users who evaluate reconstruction quality by MTF or visual sharpness alone, and it cannot be addressed because the vendor software exposes no tunable parameters. The preclinical micro-CT community would benefit from adopting task-based metrics such as NPW~$d'$ alongside traditional resolution and noise measures.

The computation time data reveal diminishing returns for iterative methods. ASTRA SIRT, at $5\times$ the computation cost of FDK, provided the best low-contrast detectability (Rose criterion at \SI{2.66}{mm} vs \SI{4.08}{mm} for FDK). TIGRE OS-SART, at $50\times$ the cost, achieved nearly identical detectability (\SI{2.76}{mm}) while exhibiting instability---its noise \emph{increased} with additional projection data. The additional $10\times$ computation of TIGRE over SIRT at default settings buys no meaningful improvement. For high-contrast tasks ($\Delta C = \SI{500}{HU}$), all methods crossed the Rose criterion below \SI{1.01}{mm} and the $\SI{0.16}{mm}$ difference between the best and worst methods does not justify a $5$--$50\times$ increase in computation time.

These results suggest practical guidance: for high-contrast preclinical imaging tasks---bone morphometry, lung airway analysis, implant visualisation---FDK is sufficient and fastest. When low-contrast detection is critical, ASTRA SIRT offers a genuine improvement at a manageable computational cost. Beyond performance, an open-source FDK pipeline with exposed parameters enables integration into workflows that vendor software cannot support, such as dual-domain deep neural network training where differentiable reconstruction operators must be embedded in the training loop, or sinogram-domain processing that requires access to intermediate data representations.

Several limitations should be noted. All quantitative evaluations were performed on a single phantom and scanner. The iterative methods were run at fixed iteration counts (100) with default parameters; systematic tuning of relaxation schedules, subset sizes, or iteration counts could shift the resolution--noise balance. Our pipeline does not currently include scatter correction or advanced ring artefact suppression, which may account for part of the noise difference relative to the vendor software. Despite these limitations, the default-settings comparison reflects what most users will encounter in practice.

%% ============================================================
\section{Conclusion}\label{sec:conclusion}
%% ============================================================

We benchmarked four cone-beam reconstruction methods on identical preclinical micro-CT data using task-based image quality metrics. The results demonstrate that reconstruction method rankings depend on the imaging task: the vendor software with the highest spatial resolution failed at low-contrast detection, while iterative methods offered genuine detectability gains but at $5$--$50\times$ the computation time. ASTRA SIRT at default settings provided the best cost--performance tradeoff among iterative methods; TIGRE OS-SART's $50\times$ computation cost yielded no additional benefit and exhibited instability across scan protocols. For the majority of preclinical micro-CT applications, FDK remains the method of choice. Our open-source implementation provides a transparent, tunable, and integrable alternative to vendor software, and is freely available to the community.

% References
\bibliographystyle{unsrtnat}
\bibliography{report}

\section*{Acknowledgments}
This work was supported by the BC Lung Foundation.

\section*{Author Contributions}
Falk L Wiegmann and Nancy L Ford contributed to the research direction and conceptualisation. Falk L Wiegmann developed the FDK pipeline, performed the benchmarking analysis, and wrote the manuscript. Nancy L Ford supervised the research, secured funding, provided critical review, and edited the manuscript.

\section*{Competing Interests}
The authors declare no competing interests.

\section*{Data Availability}
The reconstruction pipeline code is publicly available at \url{https://github.com/UBC-Ford-lab/eXplore_CT_120_fdk_reconstruction_algorithm}.
The phantom scan data used for evaluation are available from the corresponding author on reasonable request.

\end{document}